\documentclass[aps,prd,twocolumn,superscriptaddress,showpacs,showkeys]{revtex4}

\usepackage{latexsym}
\usepackage{amsmath}
\usepackage{amssymb}

\usepackage{supertabular} 
\usepackage{placeins}
\usepackage{epsfig}
\usepackage{graphicx}
\usepackage{graphics}

\usepackage{color,soul}

\usepackage{url}
\usepackage{hyperref} 
\hypersetup{
   colorlinks=true,
   linkcolor=blue,
   citecolor=blue,
   final=true,
   urlcolor=blue
}

\begin{document}
\title{LHCb pentaquarks in constituent quark models}
\author{P.G. Ortega}
\affiliation{Instituto de F\'isica Corpuscular, Universidad de Valencia, E-46071 
Valencia, Spain.}


\author{D. R. Entem}

\author{F. Fern\'andez}
\affiliation{Grupo de F\'isica Nuclear and Instituto Universitario de F\'isica 
Fundamental y Matem\'aticas (IUFFyM), Universidad de Salamanca, E-37008 
Salamanca, Spain}

\begin{abstract}
The recently discovered $P_c(4380)^+$ and $P_c(4450)^+$ states at LHCb have 
masses close to the $\bar D\Sigma_c^*$ and $\bar D^*\Sigma_c$
thresholds, respectively, which suggest that they may have  significant 
meson-baryon molecular components. We analyze these states in the framework of a 
constituent quark model which has been applied to a wide range of hadronic 
observables, being 
the model parameters, therefore, completely constrained. 

The $P_c(4380)^+$ and $P_c(4450)^+$ are studied as molecular states composed by 
charmed baryons and open charm mesons. Several bound states with the proper 
binding energy are found in the $\bar D\Sigma_c^*$ and $\bar D^*\Sigma_c$ 
channels. We discuss the possible assignments of these states from their decay 
widths. Moreover, two more states are predicted, associated with the $\bar 
D\Sigma_c$ and $\bar D^* \Sigma_c^*$ thresholds.
\end{abstract}

\maketitle

One of the most important research topics of hadron physics in the last years 
has been the hadron structure beyond the naive quark model. Already in the dawn 
of the quark models, Gell-Mann suggested~\citep{GellMann:1964nj} that, apart from 
the popular $q\bar q$ and $qqq$ configurations, there could exist multiquark 
structures.

Since 2003 plenty of new XYZ states were reported, being most of them 
candidates to multiquark configurations~\cite{brambilla2004heavy}.
Among the last XYZ states discovered, the two charm pentaquark resonances 
$P_c(4380)^+$ and $P_c(4450)^+$ were observed by the LHCb Collaboration in the 
$J/\psi p$ invariant mass spectrum in the $\Lambda^0_b\rightarrow J/\psi K^- p$ 
process~\cite{Aaij:2015tga}. The values of the masses and widths from a fit 
using Breit-Wigner amplitudes are $M_{P_c(4380)}=(4380\pm 8\pm 29)$~MeV/c$^2$, 
$\Gamma_{P_c(4380)}=(205\pm 18\pm 86)$~MeV, $M_{P_c(4450)}=(4449.8\pm 1.7\pm 
2.5)$~MeV/c$^2$ and $\Gamma_{P_c(4450)}=(39\pm 5\pm 19)$~MeV.

According to the LHCb analysis the most likely angular momentum and parity 
values
for the states are $J^P=\frac{3}{2}^{\pm}$ or $J^P=\frac{5}{2}^{\pm}$. The 
parities of the two states are opposite with the preferred spins being 
$\frac{3}{2}$ for one of the two states and $\frac{5}{2}$ for the other.

After the report of the two $P^+_c$ structures many theoretical works appeared 
suggesting different explanations, from the molecular meson-baryon pentaquark 
to kinematical triangle singularities going through diquark models or 
topological soliton models. As it is impossible, within the length of a letter, 
to cite all the 
publications we refer to the review~\cite{Chen2016}.

A common characteristic of the pentaquark structures and the XYZ states is that 
they appear in the vicinity of a two particle threshold. For example, the  
$P_c(4380)^+$ and $P_c(4450)^+$ are very close to the $\bar D\Sigma_c^*$ and 
$\bar D^*\Sigma_c$ thresholds, respectively. This fact suggests that, if there 
exist a strong enough residual interaction between 
the two particles, a bound state or a resonance can be formed. 
The structure of these bound states depends on the dynamics of the two particle 
system and this dynamics is usually model dependent. It is critical to have 
under control the strength of the residual interaction, because different 
structures can be produced depending on which threshold are involved in 
the dynamics of a potential bound state. For that reason, the interaction should be fully 
validated from the comparison against other 
experiments to avoid the generation of spurious bound states.

A model which fulfills the requirements stated above is the constituent quark 
model of Ref.~\cite{Vijande:2004he}, updated in 
Ref.~\cite{Segovia:2008zz}. The model has been extensively used to describe the 
hadron phenomenology~\cite{Valcarce:2005em,Garcilazo:2001ck,Segovia:2011dg}.

The aim of this letter is to use this model to study the possible existence of 
charm pentaquark resonances in this energy region

The most natural explanation for the two pentaquark resonance is to assume a 
 $\bar D^{(*)}\Sigma_c^{(*)}$ molecular structure, where $(*)$ denotes any combination 
 of $\bar D$ ($\Sigma_c$) or $\bar D^*$ ($\Sigma_c^*$) states. Other possible configurations like $\chi_c p$, 
which have thresholds in this energy region, are less likely due to the lack of 
residual interaction at first order between the two particles.
Taken into account that the $J^P$ of the different states are not clearly 
determined in the experiment, it would be also interesting to calculate the 
strong decays of the pentaquark resonances, which can provide guidance to the 
experimentalists.

 The constituent quark model of Ref.~\cite{Vijande:2004he} is based on the 
assumption that the light constituent mass appears due to the spontaneous 
chiral symmetry breaking of QCD at some momentum scale. Regardless of the breaking mechanism, the simplest Lagrangian 
which describe this situation must contain chiral fields to compensate the mass term and can be 
expressed as~\cite{Diakonov:2002fq}
\begin{equation}\label{lagrangian}
{\mathcal L}
=\overline{\psi }(i\, {\slash\!\!\! \partial} -M(q^{2})U^{\gamma_{5}})\,\psi 
\end{equation}
where $U^{\gamma _{5}}=\exp (i\pi ^{a}\lambda ^{a}\gamma _{5}/f_{\pi })$, 
$\pi ^{a}$ denotes nine pseudoscalar fields $(\eta _{0,}\vec{\pi }
,K_{i},\eta _{8})$ with $i=$1,...,4 and $M(q^2)$ is the constituent mass. This 
constituent quark mass, which vanishes at large momenta and is frozen at low 
momenta at a value around 300 MeV, can be explicitly obtained from the theory 
but its theoretical behavior can be simulated by parametrizing 
$M(q^{2})=m_{q}F(q^{2})$ where $m_{q}\simeq $ 300 MeV, and
\begin{equation}
F(q^{2})=\left[ \frac{{\Lambda}^{2}}{\Lambda ^{2}+q^{2}}
\right] ^{\frac{1}{2}} \, .
\end{equation} 
The cut-off $\Lambda$ fixes the
chiral symmetry breaking scale.

The Goldstone boson field matrix $U^{\gamma _{5}}$ can be expanded in terms of 
boson fields,
\begin{equation}
U^{\gamma _{5}}=1+\frac{i}{f_{\pi }}\gamma ^{5}\lambda ^{a}\pi ^{a}-\frac{1}{%
2f_{\pi }^{2}}\pi ^{a}\pi ^{a}+...
\end{equation}
The first term of the expansion generates the constituent quark mass while the
second gives rise to a one-boson exchange interaction between quarks. The
main contribution of the third term comes from the two-pion exchange which
has been simulated by means of a scalar exchange potential.

In the heavy quark sector 
chiral symmetry is explicitly broken and we do not need to introduce additional 
fields.
However the chiral fields introduced above provide a natural way to incorporate 
the pion exchange interaction in the molecular dynamics.

The other two main properties 
of QCD (besides the chiral symmetry breaking) are confinement and asymptotic freedom. 
At present it is still unfeasible to analytically derive these properties from the QCD Lagrangian, 
hence we model the interaction by 
a phenomenological confinement and the one-gluon exchange potentials, the last 
one, following De Rujula~\cite{DeRujula:1975qlm}, coming from the lagrangian.

\begin{equation}
\label{Lg}
{\mathcal L}_{gqq}=
i{\sqrt{4\pi\alpha _{s}}}\, \overline{\psi }\gamma _{\mu }G^{\mu
}_c \lambda _{c}\psi  \, ,
\end{equation}
where $\lambda _{c}$ are the SU(3) color generators and G$^{\mu }_c$ the
gluon field. 
  
The confinement term, which prevents from having colored hadrons,
can be physically interpreted in a picture where
the quark and the antiquark are linked by a one-dimensional color flux-tube.
The spontaneous creation of light-quark pairs may
give rise at same scale to a breakup of the color flux-tube. This can be 
translated into a screened potential, in such a way that the potential
saturates at the same interquark distance, such as
\begin{equation}
V_{CON}(\vec{r}_{ij})=\{-a_{c}\,(1-e^{-\mu_c\,r_{ij}})+ \Delta\}(\vec{%
\lambda^c}_{i}\cdot \vec{ \lambda^c}_{j})\,
\end{equation}
where $\Delta$ is a global constant to fit the origin of
energies. Explicit expressions for all these interactions are given in Ref.~\cite{Vijande:2004he}.
In the same reference all the parameters of the model are detailed,
additionally adapted for the heavy meson spectra in Ref.~\cite{Segovia:2008zz}.

Following Ref.~\cite{Valcarce:2005em}, in order to model the meson-baryon system we use a 
Gaussian form to describe the baryon wave function,
\begin{equation} 
\psi (\vec{p}_i)=\prod_{i=1}^3 \left[ \frac{\alpha_i b^2}{\pi}\right]^{\frac{3}{4}} 
e^{-\frac{b^2 \alpha_i p_i^2}{2}},
\end{equation}
where we take the values $b=0.518\,fm$ and $\alpha_i=1$ for the nucleon wave 
function~\cite{Valcarce:2005em}, and the scaling parameters $\alpha_i$ for different flavors are obtained using 
the prescription of Ref.~\cite{Straub:1988gj}.

In terms of Jacobi coordinates this wave function is expressed as,
\begin{equation} 
\psi =\left[ \frac{\eta b^2}{3\pi}\right]^{\frac{3}{4}}e^{-\frac{b^2\eta P^2}{6}} \phi _B 
(\vec{p}_{\xi_1},\vec{p}_{\xi_2})
\end{equation}
where $\vec{P}$ is the baryon momentum in the center of mass system and 
$\vec{p}_{\xi_1}$ and $\vec{p}_{\xi_2}$ momenta correspond to internal 
coordinates. The internal spatial wave function is written as,
\begin{equation} 
\phi _B 
(\vec{p}_{\xi_1},\vec{p}_{\xi_2})=\left[\frac{2\eta_1 b^2}{\pi}\right]^{\frac{3}{4}}e^{
-b^2\eta_1p^2_{\xi_1}}\left[\frac{3 \eta_2 b^2}{2\pi}\right]^{\frac{3}{4}}e^{-\frac{3}{4} 
b^2\eta_2p^2_{\xi_2}}
\end{equation}
To find the quark-antiquark bound states we solve the 
Schr\"odinger equation using the Gaussian Expansion 
Method~\cite{Hiyama:2003cu} with the interaction described above.

The meson-baryon interaction is derived from the $qq$ interaction by using 
the Resonating Group Method (RGM), 
introduced by Wheeler~\cite{Wheeler:1937zz} to study light nuclei but also 
widely used to study multi-quark systems~\cite{Shimizu:1989ye}.

In our case, the meson baryon interaction under evaluation has a quark 
content $\bar Qn - Qnn$, where $Q=c,b$ and $n$ are the light quarks.
Due to the presence of these light quarks, a complete interaction for this system must include a direct potential $V_D$, 
generated by $\pi$ and $\sigma$ exchanges, and an 
exchange one, $V_E$. These potentials can be expressed as
\begin{widetext}
\begin{eqnarray}
V_D(\vec{P}',\vec {P})&=&\sum_{i\in A;j\in B}\int 
\Psi^*_{l_A'm_A'}(\vec{p}_A')\Psi^*_{l_B'm_B'}(\vec{p}_B')V^D_{ij}(\vec{p}_{ij}',\vec {p}_{ij})
\Psi_{l_Am_A}(\vec{p}_A)\Psi_{l_Bm_B}(\vec{p}_B)
dp_{\xi_A}'dp_{\xi_B}'dp_{\xi_A}dp_{\xi_B}
\\
V_E(\vec{P}',\vec {P})&=&\sum_{i\in A,j\in B}\int 
\Psi^*_{l_A'm_A'}(\vec{p}_A')\Psi^*_{l_B'm_B'}(\vec{p}_B')V^E_{ij}(\vec{p}_{ij}',\vec
{p}_{ij})\Psi_{l_Am_A}(\vec{p}_A)\Psi_{l_Bm_B}(\vec{p}_B)
dp_{\xi_A}'dp_{\xi_B}'dp_{\xi_A}dp_{\xi_B}
\end{eqnarray}
\end{widetext}
which gives the residual interaction between clusters and, at the same time,
describes the strong decays of the potential bound states into 
the different channels like $\bar D^{(*)}\Lambda_c$, with direct potentials, 
or $J/\psi N$, done by simple quark rearrangement driven by the quark interaction.


Exploiting the symmetries of the system there are six possible diagrams which 
contribute to this coupling.
The $h_{fi}$ matrix elements corresponding to each diagram is the product of 
three factors
\begin{widetext}
\begin{eqnarray} 
	h_{ij}(\vec P',\vec P) &= S& 
	\langle \phi_{\bar D^{(*)}} \phi_{\Sigma_c^{(*)}}|H_{ij}^O| \phi_{\bar D^{(*)}} \phi_{\Sigma_c^{(*)}} \rangle
	\langle \xi_{\bar D^{(*)}\Sigma_c^{(*)}}^{SFC} |\mathcal{O}_{ij}^{SFC}| 
	\xi_{\bar D^{(*)}\Sigma_c^{(*)}}^{SFC} \rangle
\end{eqnarray}
\end{widetext}
where $S$ is a phase characteristic of each diagram, resulting from the 
permutation between fermion operators. This potential involves 
the same interquark interactions as the direct potentials, that is, 
$\pi$ and $\sigma$ interactions, plus contributions 
of both, the OGE and confinement potentials. 

The coupled channel equations are solved through the Lippmann-Schwinger equation 
for the t matrix
\begin{widetext}
\begin{equation}
 t^{\beta\beta'}( p, p',E)=V^{\beta\beta'}_T( p, p',E)-\sum_{\beta''}\int dq q^2 
\frac{V^{\beta\beta''}_T( p, q,E)t^{\beta''\beta'}( q, p',E)}{q^2/(2\mu)-E-i0}
\end{equation}
\end{widetext}
where $\beta$ specifies the quantum numbers necessary to define a partial wave 
in the baryon meson state.
Finding the poles of the $t(\vec p,\vec p',E)$ matrix 
we will determine the mass and the quantum numbers of 
the  molecules.

The decay of the particle is calculated through the standard formula
\begin{equation} \label{ec:decayL2940}
 \Gamma = 2\pi \frac{E_AE_Bk_0}{M_{P_c}} \sum_{J,L}|\mathcal{M}_{J,L}|^2
\end{equation}
where $E_A$ and $E_B$ are the relativistic energies of the final state hadrons 
$\bar D^{(*)}\Lambda_c$ or $J/\psi N$, $M_{P_c}$ is the mass of the pentaquark and $k_0$ is the 
on-shell momentum of the system, given by,
\begin{equation}
k_0=\frac{\sqrt{[M_{P_c}^2-(M_A-M_B)^2][M_{P_c}^2-(M_A+M_B)^2]}}{2M_{P_c}}.
\end{equation}

To calculate the final amplitude of the process $\mathcal{M}$ the wave function 
of the molecular state is used,
\begin{equation} \label{ec:amplitudL2940}
 \mathcal{M}=\int_0^\infty V_{\bar D^{(*)}\Sigma_c\to AB}(k_0,P)\chi_{\bar D^{(*)}\Sigma_c}(P)\,P^2dP
\end{equation}
where $V_{\bar D^{(*)}\Sigma_c\to AB}(k_0,P)$ is the potential to the final state and 
$\chi_{\bar D^{(*)}\Sigma_c}$ is the pentaquark wave function.


Exploring the most interesting channels for the $\bar D^{(*)}\Sigma_c^{(*)}$ we obtain the pentaquark
candidates shown in Table~\ref{t1}.

\begin{center}
\begin{table}
\begin{tabular}{ccccc}
\hline
\hline
Molecule & $J^P$ & $I$ & $Mass (MeV/c^2)$& $B_E (MeV/c^2)$ \\
\hline

$\bar D\Sigma_c$ & $\frac{1}{2}^-$ & $\frac{1}{2}$ & 4320.782& 0.765 \\
\hline
$\bar D\Sigma_c^*$ & $\frac{3}{2}^-$ & $\frac{1}{2}$ & 4384.993 & 0.993 \\
\hline
$\bar D^*\Sigma_c$ & $\frac{1}{2}^-$ & $\frac{1}{2}$ & 4458.894 & 3.796 \\
$\bar D^*\Sigma_c$ & $\frac{3}{2}^-$ & $\frac{1}{2}$ & 4461.284 & 1.406 \\
$\bar D^*\Sigma_c$ & $\frac{3}{2}^+$ & $\frac{1}{2}$ & 4462.677 & 0.013 \\
\hline
$\bar D^*\Sigma_c^*$ & $\frac{1}{2}^-$ & $\frac{1}{2}$ & 4519.792 & 7.338 \\
$\bar D^*\Sigma_c^*$ & $\frac{3}{2}^-$ & $\frac{1}{2}$ & 4523.275 & 3.855 \\
$\bar D^*\Sigma_c^*$ & $\frac{5}{2}^-$ & $\frac{1}{2}$ & 4524.552 & 2.578 \\
$\bar D^*\Sigma_c^*$ & $\frac{5}{2}^+$ & $\frac{1}{2}$ & 4526.165 & 0.965 \\
\hline
\hline
\end{tabular}
\caption{\label{t1} Masses of the different molecular states}
\end{table}
\end{center}
We consider the $\bar D^{(*)}\Sigma^{(*)}$ thresholds, which are the only ones
where a sizable residual interaction can be expected, 
mainly due to the pion exchanges. As stated above, other structures like $\chi_{c1} p$ do not 
have, in our model, residual interaction at first order and, hence, it is unlikely 
that they can develop a pentaquark structure.
In the mass region of the $P_c(4380)^+$ we obtain one $\bar D\Sigma^*_c$ state 
with $J^P=\frac{3}{2}^-$. Its mass is very close to the experimental one (note 
that the calculation performed to obtain this values is parameter free) and should, in 
principle, be identified with $P_c(4380)^+$.

Referring to the channel $\bar D^*\Sigma_c$ we found three almost-degenerated states 
around M=4460 MeV/$c^2$ with $J^P=\frac{1}{2}^-$,
$\frac{3}{2}^-$ and $\frac{3}{2}^+$. The existence of these three degenerated 
states may be the origin of the uncertainty in the experimental value of $J^P$.
The energy of those states makes them natural candidates for the $P_c(4450)^+$.

\begin{center}
\begin{table}
\begin{tabular}{ccccc}
\hline
\hline
Molecule & $J^P$ & $I$ & Width $J/\psi p$ & Width $\bar D^*\Lambda_c$ \\
\hline
$\bar D\Sigma_c$ & $\frac{1}{2}^-$ & $\frac{1}{2}$ & 2.394 & 1.109 \\
\hline
$\bar D\Sigma_c^*$ & $\frac{3}{2}^-$ & $\frac{1}{2}$ & 10.046 & 14.688 \\
\hline
$\bar D^*\Sigma_c$ & $\frac{1}{2}^-$ & $\frac{1}{2}$ & 5.294 & 63.576 \\
$\bar D^*\Sigma_c$ & $\frac{3}{2}^-$ & $\frac{1}{2}$ & 0.794 & 21.198 \\
$\bar D^*\Sigma_c$ & $\frac{3}{2}^+$ & $\frac{1}{2}$ & 0.214 & 6.292 \\
\hline
$\bar D^*\Sigma_c^*$ & $\frac{1}{2}^-$ & $\frac{1}{2}$ & 0.893 & 9.954 \\
$\bar D^*\Sigma_c^*$ & $\frac{3}{2}^-$ & $\frac{1}{2}$ & 22.901 & 4.050 \\
$\bar D^*\Sigma_c^*$ & $\frac{5}{2}^-$ & $\frac{1}{2}$ & 0.053 & 3.048 \\
$\bar D^*\Sigma_c^*$ & $\frac{5}{2}^+$ & $\frac{1}{2}$ & 0.051 & 0.845 \\
\hline
\hline
\end{tabular}
\caption{\label{t2} Widths, in MeV, of the different molecular states}
\end{table}
\end{center}

Finally, if we look to the $\bar D\Sigma_c$ and $\bar D^*\Sigma_c^*$ channels, 
we found one state in the first channel with $J^P=\frac{1}{2}^-$
and four almost-degenerated states around 4523 MeV/$c^2$ with $J^P=\frac{1}{2}^-$, 
$\frac{3}{2}^-$, $\frac{5}{2}^-$ and $\frac{5}{2}^+$. The first state is around 
60 MeV/$c^2$ lower than the one found in the $\bar D\Sigma_c^*$ channel but with 
different quantum numbers. The second four states are higher in energy than
the $P_c(4450)^+$. Both may correspond to new pentaquark states.

In order to obtain a deeper insight into the structure of the pentaquarks we have 
studied the
decay channels $J/\psi p$, the channel in which the resonances were 
discovered, and $\bar D^*\Lambda_c$. The corresponding widths for both channels are 
shown in Table~\ref{t2}.

The first observation that can be made from these results is that the decay width 
through the  $\bar D^*\Lambda_c$ channel is generally equal to or greater than the 
width via the  $J/\psi p$ channel. This suggests that the $\bar D^*\Lambda_c$ 
channel is a suitable channel for studying the properties of these resonances. 
In particular, the width of the predicted $\bar D^*\Sigma_c$ resonance with 
$J^P=\frac{1}{2}^-$ is twelve times greater through the $\bar D^*\Lambda_c$ channel
than through the $J/\psi p$ channel, being this decay a good check for the 
existence of the resonance.

The second observation is that the width of the $\bar D\Sigma_c^*$ 
$J^P=\frac{3}{2}^-$ state is too small to explain the experimental one, whereas 
the values of the widths in the $\bar D^*\Sigma_c$ are more compatibles with the 
experiment.

Concerning the parity of the states,  a molecular scenario is not 
the most convenient to obtain positive parity states 
because, being the $\bar D^{(*)}$ mesons and the $\Sigma_c^{(*)}$ baryons of 
opposite parity, the relative angular momentum should be at least $L=1$ 
(P-wave) which will be above S-waves.
This is reflected in the fact that the states with 
positive parity in Table~\ref{t1} are those with smaller binding energies.

The authors of Ref.~\cite{Roca:2016tdh} argued that, using the spin suggested by 
the experimental analysis, the most likely assignment for spin parity of 
both pentaquarks are $J^P=(\frac{3}{2}^-,\frac{3}{2}^-)$ or 
$(\frac{3}{2}^-,\frac{5}{2}^+)$ and much less likely 
$(\frac{5}{2}^+,\frac{3}{2}^-)$. The first combination is present in our results 
although based on the decay widths our favorite combination would be 
$(\frac{3}{2}^-,\frac{1}{2}^-)$.

Although the two pentaquark states decaying to $J/\Psi p$ should have $I=\frac 1 2$, one could
consider the possibility of $I=\frac 3 2$ pentaquarks decaying to $J/\Psi N\pi$ through a $\Delta$.
We have investigated this possibility and we did not find any such state.

Let us now compare our results with those of some other molecular models available on 
the literature. Roca {\it et al.}~\cite{Roca:2015dva}, using  a coupled-channel 
unitary approach within the local hidden gauge formalism, found that the 
$P_c(4450)^+$ is a $\bar D^*\Sigma_c$-$\bar D^*\Sigma_c^*$ molecular state with 
$I=\frac{1}{2}$ and $J^P=\frac{3}{2}^-$. Although it seems similar to our 
result, a careful analysis shows that the binding energies predicted by this 
model are on the order of 45 MeV/c$^2$, whereas in our case the binding energies are always less than 10
MeV/$c^2$. This is the reason why a second $\bar D^*\Sigma_c^*$ component appears 
in Ref.~\cite{Roca:2015dva}. In any case these differences are relevant to 
discriminate between the two models.

Using a model of meson exchanges combined with a Bethe-Salpeter equation, 
He~\cite{He:2015cea} investigated different molecular channels. As in our case,  
He obtained a bound state with $J^P=\frac{3}{2}^-$ spin from the $\bar D 
\Sigma_c^*$ interaction, consistent with the $P_c(4380)^+$. From the $\bar D^*\Sigma_c$ channel a bound state with 
$J^P=\frac{5}{2}^+$ is produced, which can be related to 
the $P_c(4450)^+$. However, in order to obtain this last state, one has to move the cut-off 
from 1 GeV to almost 3 GeV.

Moreover, Chen {\it et al.}~\cite{Chen:2016heh} obtained similar results to those 
of Ref.~\cite{Roca:2015dva} in the framework of an OPE model, finding a $\bar 
D^*\Sigma_c$ molecular state with ($I=\frac 1 2$, $J^P=\frac 3 2^-$) quantum numbers and a $\bar 
D^*\Sigma_c^*$ molecular state with ($I=\frac 1 2$, $J^P=\frac 5 2^-$) in the same mass range 
 that the observed $P_c(4380)^+$ and $P_c(4450)^+$ respectively. Again, the model 
should predict a strong residual interaction in order to lower the respective thresholds 
to the physical masses.

As a summary, our results confirm the fact that there are several states with a
$\bar D^{(*)}\Sigma_c^{(*)}$ structure 
in the vicinity of the masses of the $P_c(4380)^+$ and $P_c(4450)^+$ pentaquark
states reported by the LHCb. However, more theoretical and experimental work is 
needed to completely clarify the nature of these states.

\acknowledgments
This work has been partially funded by MINECO
under Contract
No. FPA2013-47433.C2-2-P, by the Spanish Excellence Network on Hadronic
Physics FIS2014-57026-REDT and
by FPA2015-69037-REDC.

\bibliographystyle{apsrev}
\bibliography{pentaquark}

\end{document}